\documentclass[aps,a4paper,twocolumn,nofootinbib]{revtex4}

\usepackage{amsmath}
\usepackage{graphicx}
\usepackage{subfigure}
\usepackage[usenames,dvipsnames]{color}
\definecolor{darkblue}{RGB}{0,0,196}
\usepackage[colorlinks=true,linkcolor=darkblue,citecolor=darkblue,urlcolor=darkblue]{hyperref}
\usepackage{setspace}
\usepackage{multirow}
\usepackage{hyperref}
\usepackage{xcolor}
\usepackage{xspace}
\hypersetup{
  colorlinks   = true, 
  urlcolor     = blue, 
  linkcolor    = blue, 
  citecolor   = blue 
}
\usepackage{graphicx}
\usepackage{amsmath,bbm}
\usepackage{amssymb,bm}
\usepackage{yfonts}
\usepackage{comment}
\usepackage[normalem]{ulem}

\newcommand{\dphi}{\ensuremath{\Delta \varphi}\xspace}
\newcommand{\dy}{\ensuremath{\Delta y}\xspace}
\newcommand{\deta}{\ensuremath{\Delta \eta}\xspace}
\newcommand{\pT}{\ensuremath{p_{\rm T}}\xspace}
\newcommand{\pythia}{\textsc{PYTHIA}\xspace}
\newcommand{\gevc[1]}{\ensuremath{#1\,\text{GeV}/c}\xspace} 
\newcommand{\ddbar}{\ensuremath{D^0\bar{D}^0}\xspace}
\newcommand{\dd}{\ensuremath{D^0D^0}\xspace}
\newcommand{\dbardbar}{\ensuremath{\bar{D}^0\bar{D}^0}\xspace}
\newcommand{\ccbar}{\ensuremath{c\bar{c}}\xspace}
\newcommand{\bbbar}{\ensuremath{b\bar{b}}\xspace}
\newcommand{\pp}{pp\xspace}

\newcommand{\pbpb}{Pb--Pb\xspace}
\newcommand{\jpsi}{\ensuremath{J/\psi}\xspace}
\newcommand{\sppt}[1]{\ensuremath{\sqrt{s} = #1\,\text{TeV}}\xspace}

\begin{document}

\title{How to isolate charm production mechanisms using charm-anticharm rapidity and azimuthal angle differences}
\author{Marcus O'Riordan\footnote{ma0721or-s@student.lu.se}}
\author{Peter Christiansen\footnote{peter.christiansen@fysik.lu.se}}

\affiliation{Division of Particle and Nuclear Physics, Department of Physics, Lund University, Lund, 22364, Sweden}

\date{\today}

\begin{abstract}
Charm production in hadronic collisions proceeds via many different processes, which makes it hard to calculate from first principles. Using \pythia simulations, we study correlations between charm and anticharm hadrons and show that hard-process charm, final-state splitting charm, and initial-state-shower charm populate markedly different regions of rapidity and azimuthal-angle difference. In particular, selections on the pair rapidity separation, \dy, can strongly enrich or suppress the near-side final-state splitting contribution and thereby provide experimental sensitivity to the different production processes. A comparison with existing LHCb \ddbar correlation measurements from Run 1 motivates more differential studies using the substantially larger LHC data sets now available. Extending such measurements to heavy-ion collisions could further distinguish medium modifications of charm produced directly in the hard scattering from charm originating from a parent gluon, providing a new handle on the different interactions of heavy quarks and gluons with the quark-gluon plasma.
\end{abstract}
  
\maketitle


\section{Introduction}
\label{intro}

Charm hadrons are both important tools for testing QCD in proton-proton collisions~\cite{Frixione:1997ma,Andronic:2015wma} and a powerful probe of the hot and dense system produced in heavy-ion collisions~\cite{Andronic:2015wma}. In heavy-ion collisions, charm production at low \pT probes thermalization and the buildup of flow, while charm production at high \pT probes the medium via jet quenching. Charm is a heavy quark, meaning that its production can in principle be calculated perturbatively. However, when one looks at comparisons between measured spectra of charm mesons in proton-proton collisions and state-of-the-art calculations using perturbative QCD, it is not uncommon to find uncertainties of order 50\% for $\pT \approx \gevc[5]$~\cite{ALICE:2023sgl}. This also means that, even when measurements can be experimentally controlled using binary-collision scaling from pp to heavy-ion collisions, the theoretical input has large uncertainties that limit the physics that can be extracted. In fact, many of the ideas related to these correlation functions in heavy-ion collisions, e.g., Refs.~\cite{Zhao:2025cnp,Zhu:2006er,Parida:2023qju,Sheibani:2025iuh}, have inspired the present work.

In this paper, we propose to study correlations between charm and anticharm hadrons to separate samples enriched in different charm-production topologies. Similar approaches to investigating heavy-quark production mechanisms using angular correlations have been studied at Tevatron energies~\cite{Field:2002da}, at RHIC~\cite{Mischke:2008af} for heavy-ion collisions, and at the LHC~\cite{ATLAS:2016anw}. Related \pythia studies have also explored the separation of charm-production mechanisms using event-shape selections~\cite{Horvath:2023lho}. Here, we focus on how samples enriched in different charm-production topologies can be constructed using selections on both azimuthal angle, \dphi, and rapidity separation, \dy. The aim is to provide new experimental guidance that can help constrain the interpretation and theoretical description of charm production in pp collisions, thereby also improving the understanding of \pbpb results. In the following we use an operational classification based on the \pythia event record rather than identifying the categories one-to-one with the conventional perturbative labels of flavor creation, flavor excitation, and gluon splitting~\cite{Field:2002da,Andronic:2015wma}. We distinguish charm produced in the hard process, charm produced through final-state gluon splitting, and charm whose ancestry reaches the initial-state shower as these event-history categories lead to markedly different pair correlations. For example, final-state splitting gives charm pairs that are close in both rapidity and azimuthal angle, whereas charm produced in the hard process is predominantly back-to-back and has a much broader rapidity separation. 
This makes it possible to select regions of phase space that are strongly enriched in different production topologies, which is the central idea explored in this paper. 

The paper is organized in the following way. The simulations and the classification of events are discussed in Sec.~\ref{sec:pythia}. In Sec.~\ref{sec:lhcb}, the simulated correlation functions are compared with LHCb measurements and we demonstrate how selections in rapidity difference can construct samples enriched in the underlying production topologies. The discussion section, Sec.~\ref{sec:discussion}, focuses on how one can further improve the analysis and what the implications could be for \pbpb collisions. Finally, Sec.~\ref{sec:conclusions} reports the conclusions.

\section{PYTHIA Simulations}
\label{sec:pythia}

The simulations are performed using \pythia version 8.303 with the default tune~\cite{Bierlich:2022pfr}. Since we want to compare with LHCb measurements of prompt \ddbar correlations, each event is required to have two $D$ mesons in the LHCb acceptance, $2 < y < 4$. A total of approximately 1.2 million simulated inelastic \pp collision at \sppt{7} fulfilling this requirement were retained (out of approximately 2.3 billion events). In the \pythia simulations, weak decays were disabled, so that $D$ mesons are either directly produced or from strong or electromagnetic decays of excited $D^*$ mesons, i.e., no $D$ mesons from $B$ decays. To classify the data, \pythia's mother function is used iteratively on each charm meson to trace the charm-quark ancestry through the event record. Based on where this ancestry is found in the \pythia event history, we define three operational categories for the charm pair originating from the same production history:
\begin{itemize}
    \item \emph{Hard-process charm}: charm produced through flavor creation in a hard subprocess, $gg \rightarrow \ccbar$ or $q\bar{q} \rightarrow \ccbar$.
    \item \emph{Final-state splitting charm}: charm produced through $g \rightarrow \ccbar$ from an outgoing gluon associated with a hard process or from final-state shower evolution. This includes charm splitting from any gluon that is radiated from a final-state parton. 
    \item \emph{Initial-state-shower charm}: charm whose traced ancestry reaches the spacelike initial-state shower before the hard scattering. This category includes the conventional flavor-excitation topology, in which an incoming charm quark participates in a hard scattering, e.g. $cg \rightarrow cg$, and its ancestry is traced through backward shower evolution to a $g \rightarrow \ccbar$ splitting. This includes charm splitting from any gluon that is radiated from an initial-state parton. 
\end{itemize}
These labels describe where the charm appears in the \pythia event history and should not be interpreted as a unique decomposition of QCD amplitudes. In the simulation, the final-state splitting category is strongly concentrated on the near side, while hard-process charm is predominantly back-to-back. The initial-state-shower category is also predominantly away-side, but its azimuthal correlation is substantially broader. Experimentally, one will also associate pairs for which the two charm hadrons originate from different \ccbar production histories, which therefore forms a fourth uncorrelated category when comparing with experiment. These uncorrelated pairs include pairs from separate multiparton interactions as well as pairs from different parts of a shower.

There are at least two more categories that we did not include as the observed rates were negligible ($\ll 0.1\%$): production related to the hadronization of the beam remnants and associated production of heavy quarks with a $W$ boson.

The simulation will be compared with measurements of \ddbar correlations by LHCb~\cite{LHCb:2012aiv} in the next section. The available LHCb results include correlations, $C_2(\dy) = (1/N_{\text{pairs}})dN_{\text{pairs}}/d\dy$ and $C_2(\dphi)$, as a function of the rapidity distance, $\dy = y_1 - y_2$, and the azimuthal angle difference, $\dphi = \varphi_1 - \varphi_2$~\footnote{Note that the correlation functions presented here are always normalized so that the integral over the shown range is one.}. The corresponding \pT spectrum, $(1/N)dN/d\pT$, is also available for the events with \ddbar pairs.  Since the LHCb results were not corrected for the pair acceptance using event mixing, the $|\dy|$ distribution retains a strong acceptance effect. For uncorrelated particles distributed approximately uniformly within $2<y<4$, this leads to an approximately triangular distribution, decreasing towards zero at $|\dy|=2$.  

\begin{figure*}
\centering
\includegraphics[width=0.3\linewidth]{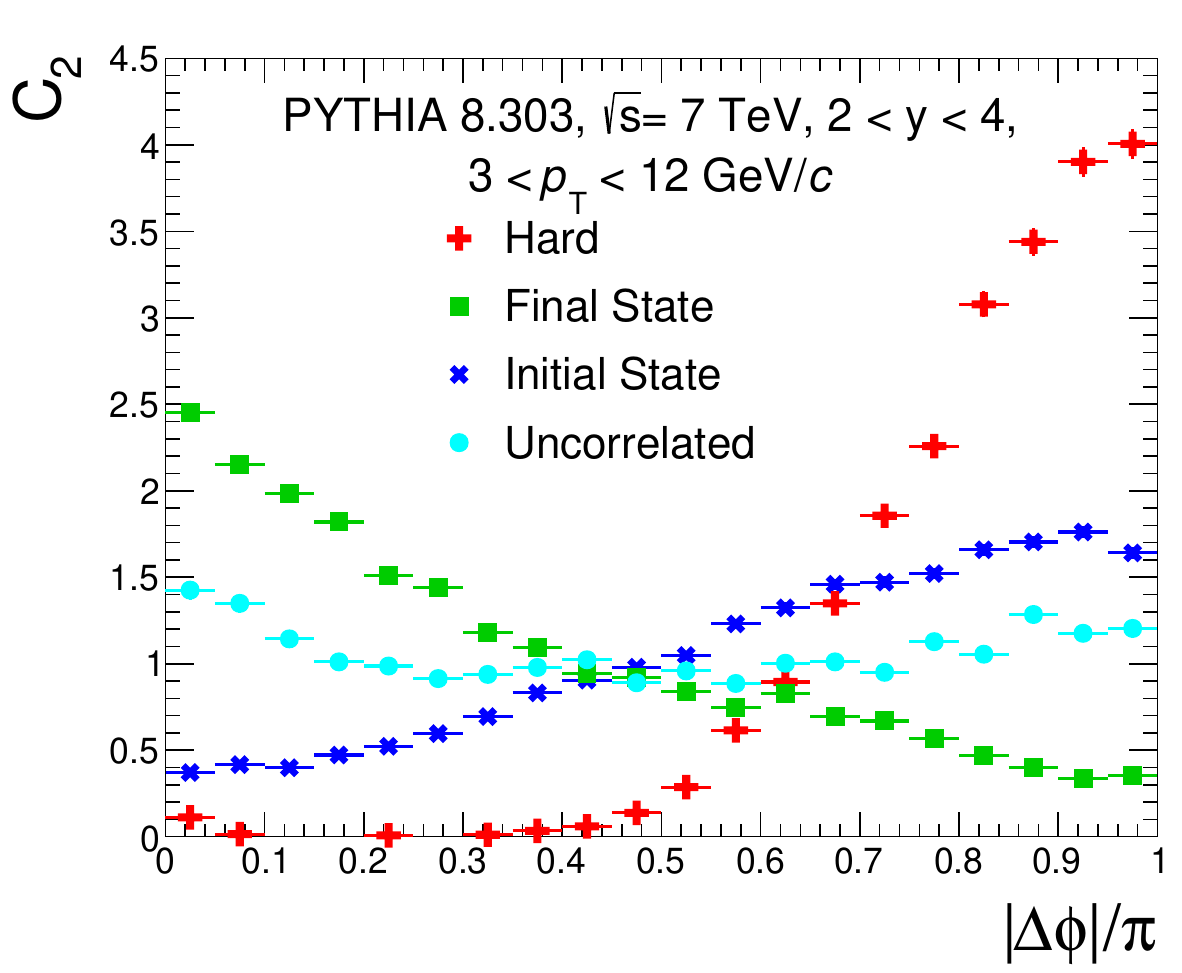}
\includegraphics[width=0.3\linewidth]{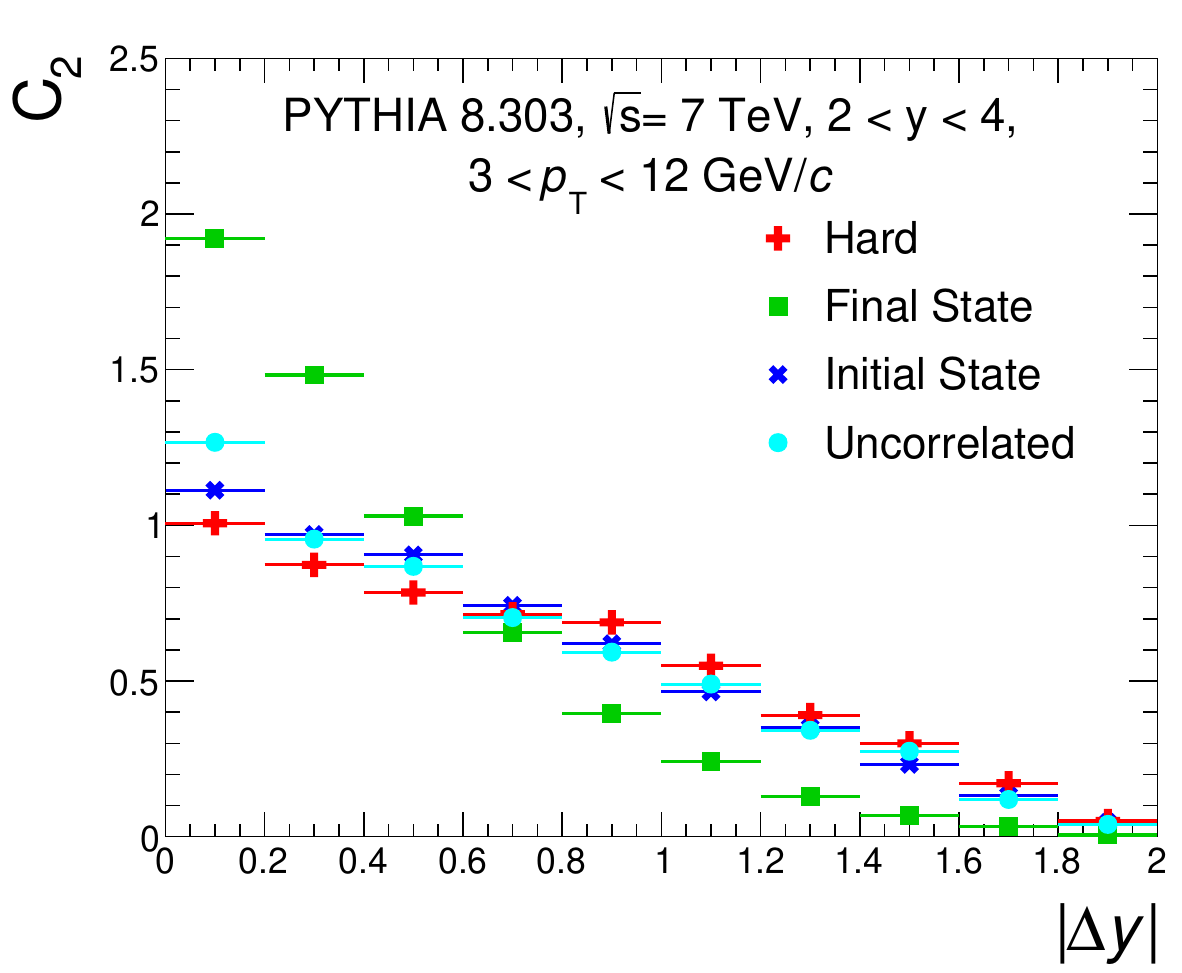}
\includegraphics[width=0.3\linewidth]{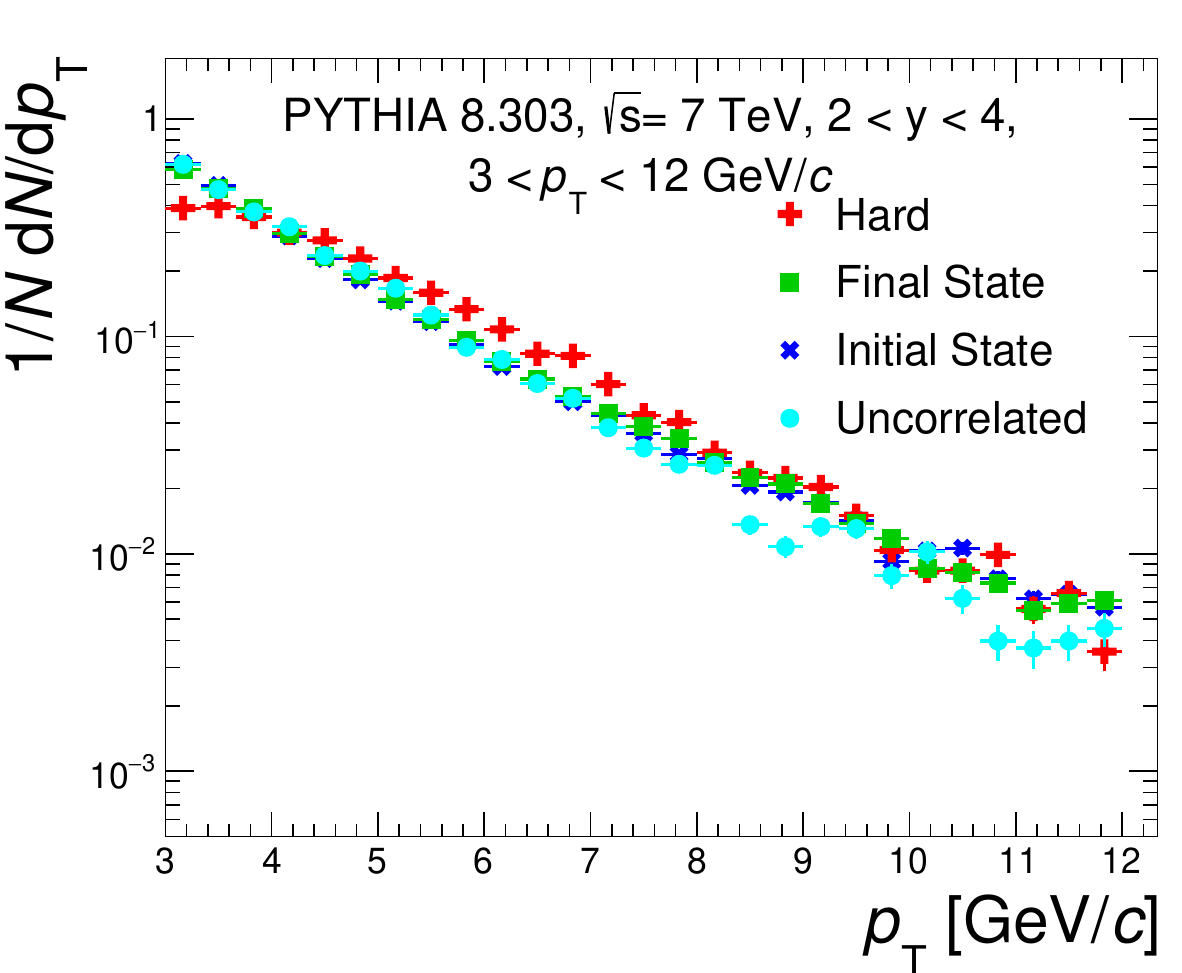}
    \caption{The distributions obtained with \pythia for the three \ddbar processes and also for uncorrelated pairs. The distributions have all been normalized to have the same integral (unity) so that the shapes can easily be compared. Left: \dphi correlation functions. Center: \dy correlation functions. Right: \pT spectra.}
\label{fig:shape}
\end{figure*}

Figure~\ref{fig:shape} compares the distributions for the three production topologies obtained from \pythia, together with the uncorrelated contribution. Apart from the \pT distributions, the three process shapes are quite distinct. Hard-process charm has a clear back-to-back peak in \dphi and an almost linear decrease in \dy, indicative of a broad rapidity separation within the LHCb acceptance. In contrast, final-state splitting charm has a strong near-side peak and a steep falloff in \dy, corresponding to a concentration near $\dy \sim 0$. Initial-state-shower charm is predominantly away-side like the hard-process category, but the azimuthal correlation is much broader. Finally, the uncorrelated component is, as expected, approximately flat in \dphi and decreases approximately linearly in \dy because of the acceptance. 

\section{Comparisons with LHCb data}
\label{sec:lhcb}

In this section we want to compare the simulation results obtained in the previous section with the measured LHCb results~\cite{LHCb:2012aiv}. We first show the direct comparison with \pythia and then investigate how the different event-history components can be constrained experimentally.

\begin{figure*}
\centering
\includegraphics[width=0.3\linewidth]{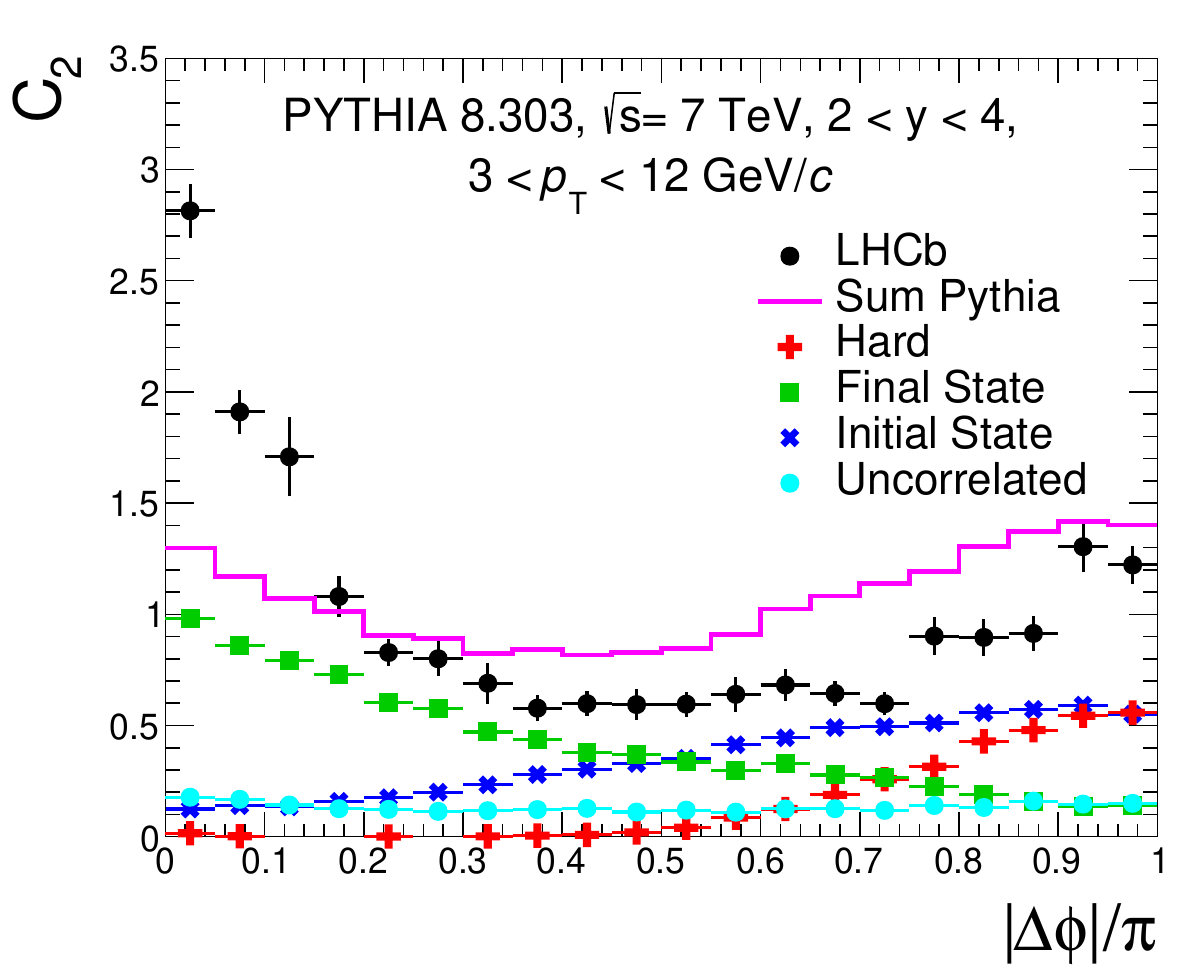}
\includegraphics[width=0.3\linewidth]{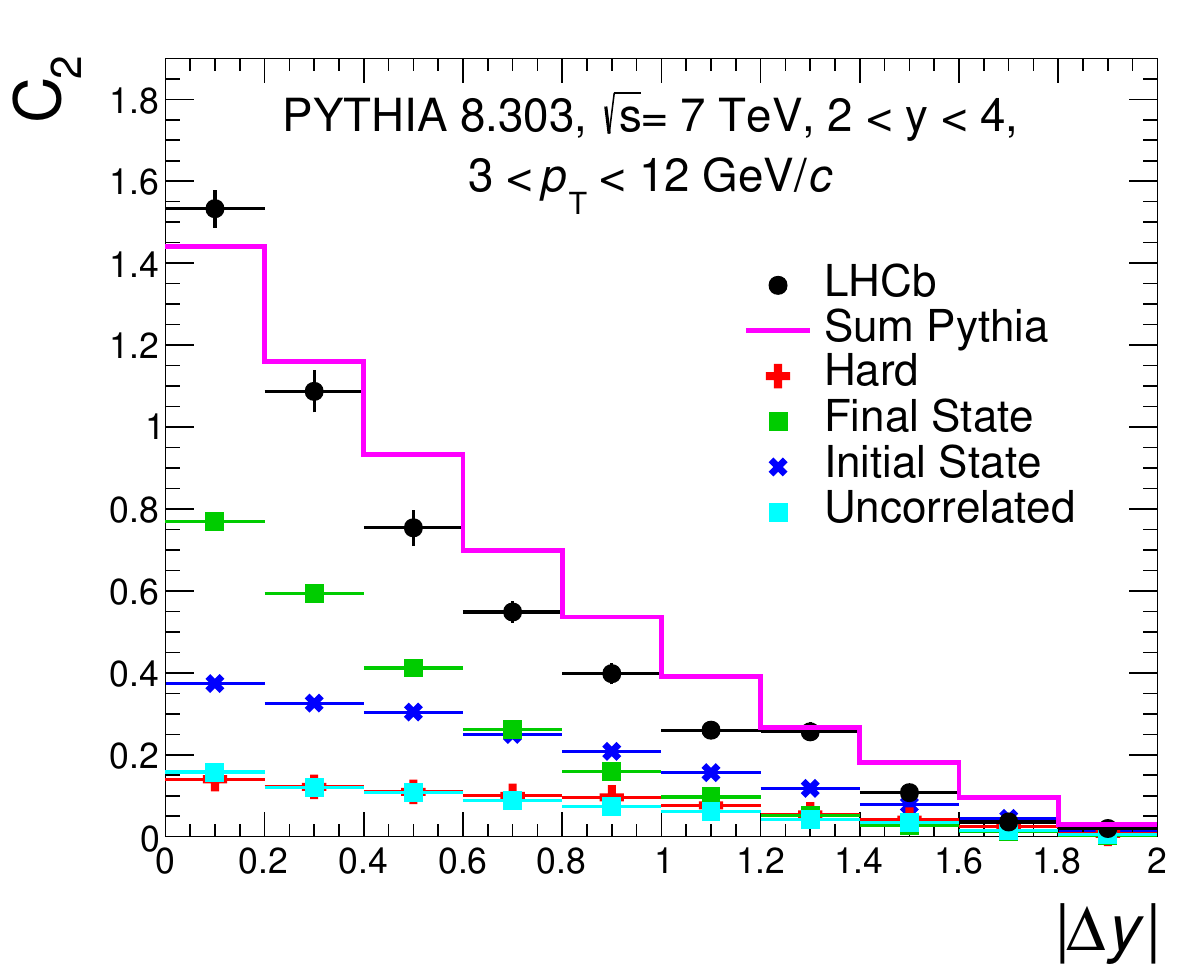}
\includegraphics[width=0.3\linewidth]{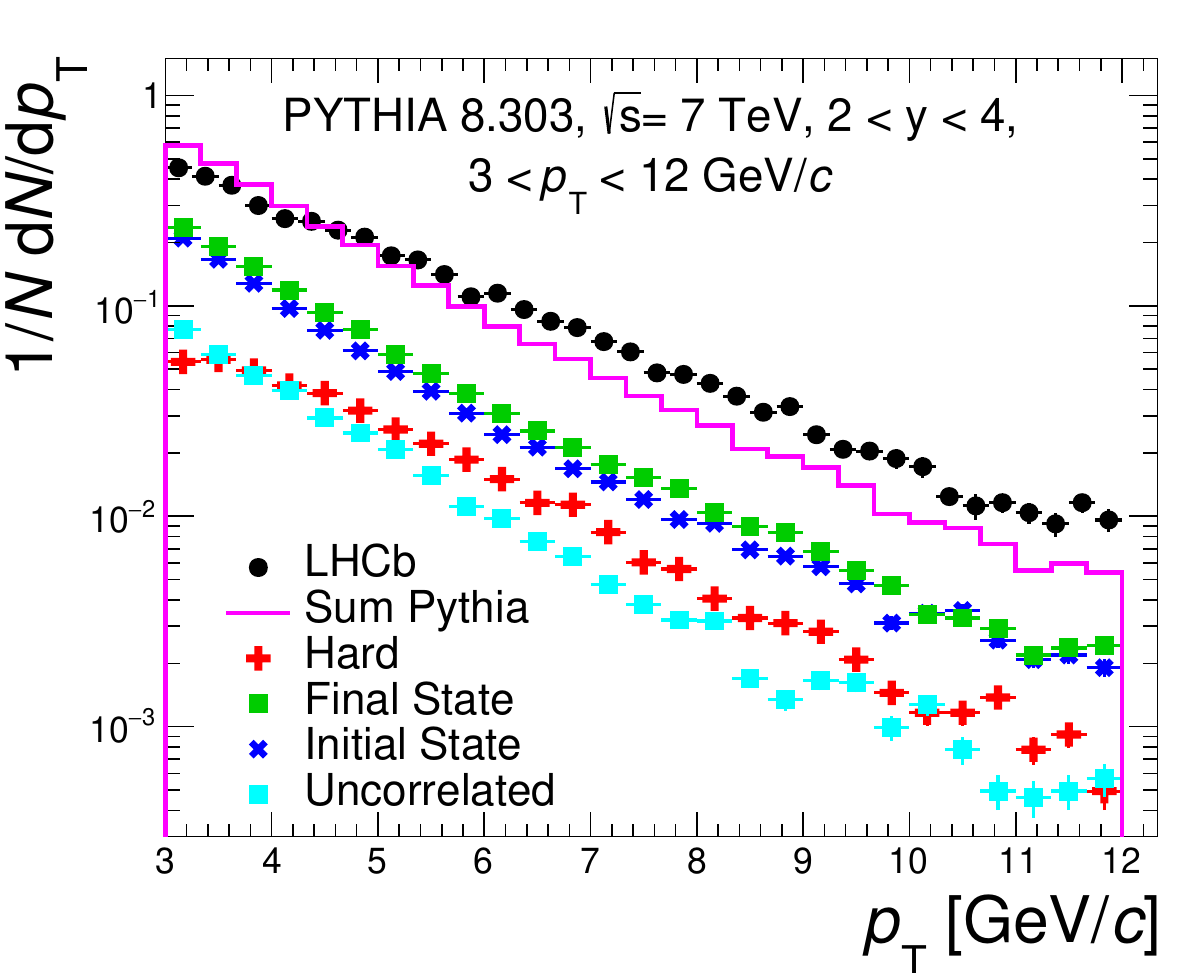}
    \caption{Comparison of LHCb results (black markers) with the summed \pythia predictions (pink lines). For the \pythia simulation, each of the four components are also shown separately. Left: \dphi correlation functions. Center: \dy correlation functions. Right: \pT spectra.}
\label{fig:pythia}
\end{figure*}

Figure~\ref{fig:pythia} shows the out-of-the-box comparison of LHCb data with \pythia simulations. As can be seen, the data have much stronger peaks at $\dphi \sim 0$ and $\dy \sim 0$, indicating that the near-side contribution associated in our classification with final-state splitting charm is underestimated by \pythia. The \pT spectra are in general also too soft in \pythia. 

\begin{figure}
\centering
\includegraphics[width=0.9\linewidth]{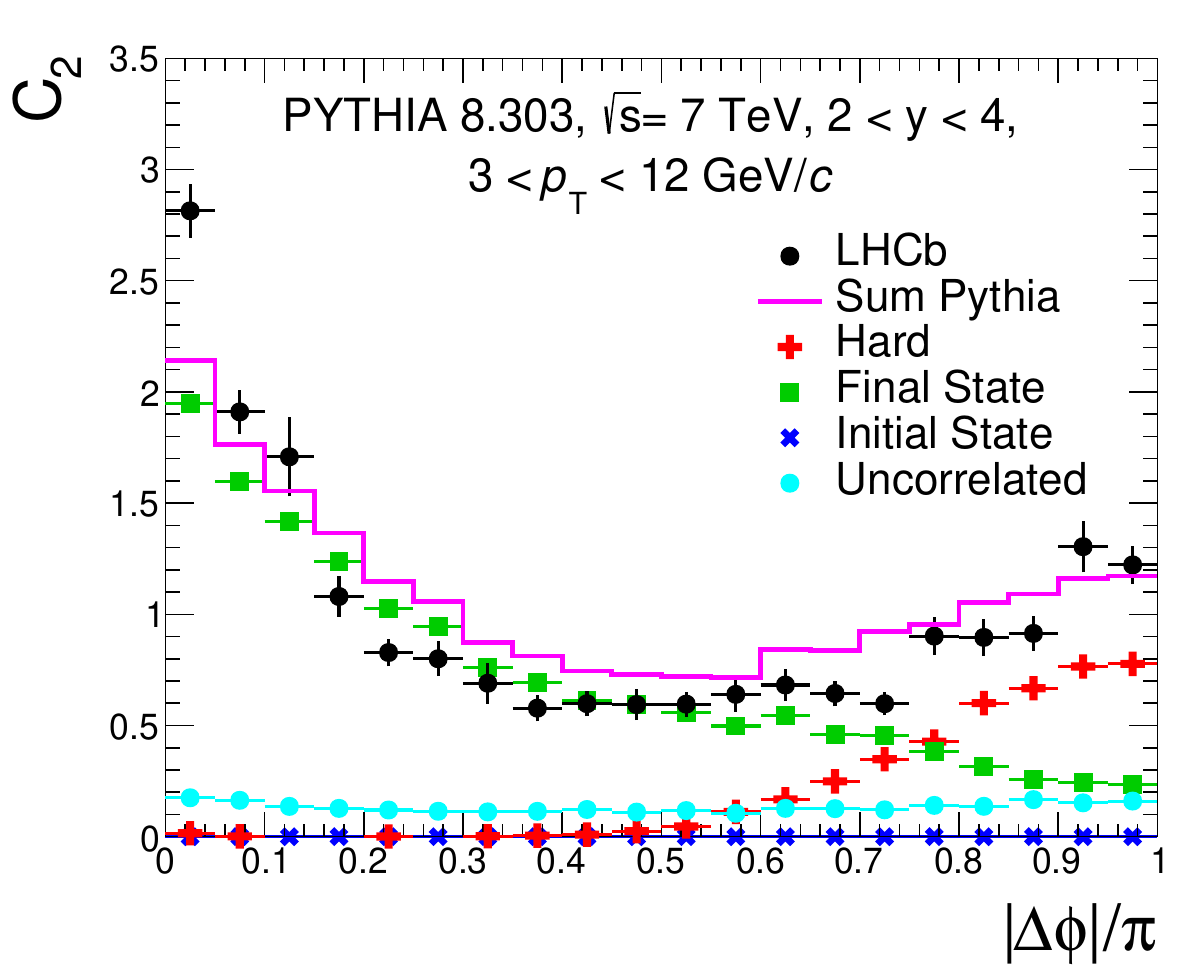}
\includegraphics[width=0.9\linewidth]{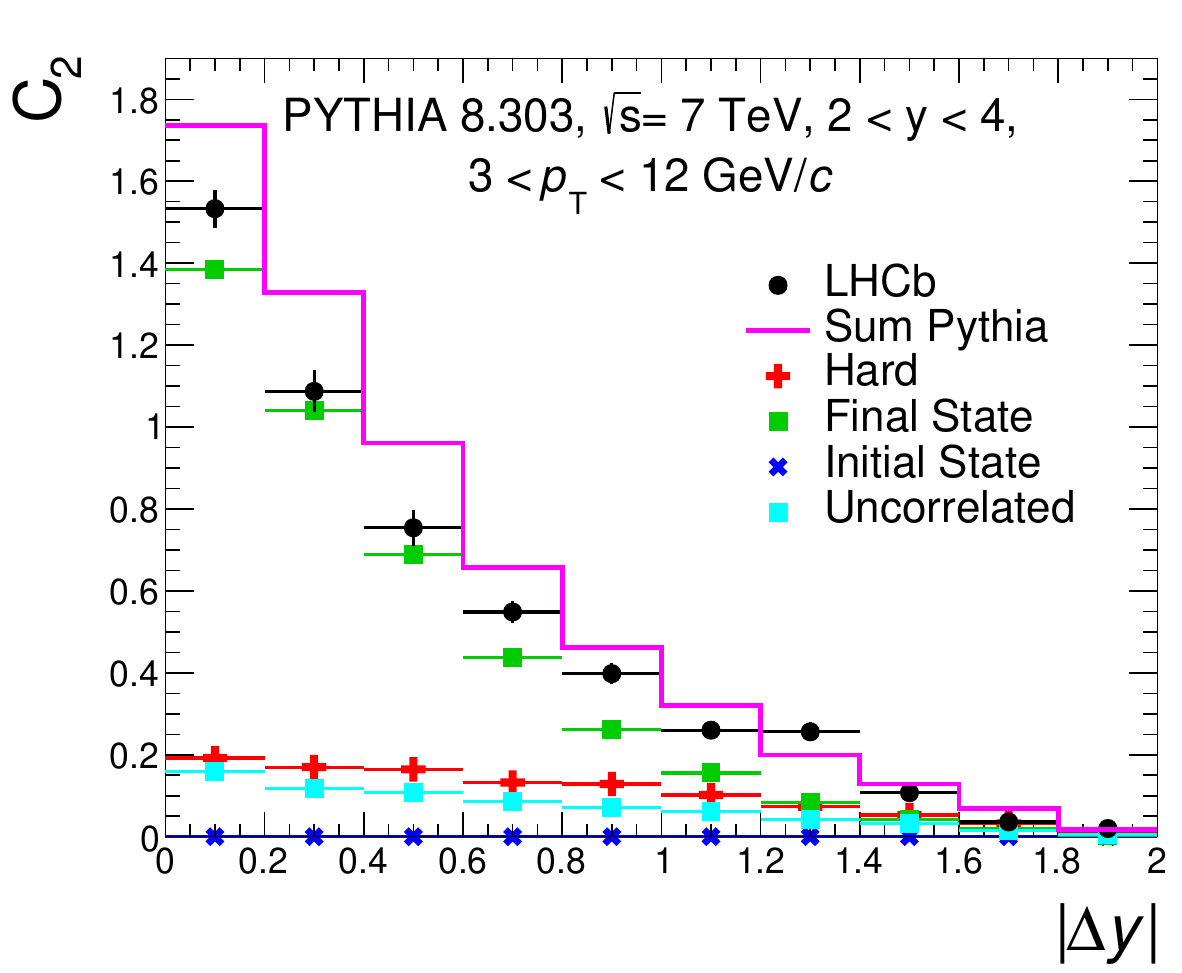}
    \caption{Comparison of LHCb results (black markers) with \emph{fitted} \pythia simulations (pink lines). The four components with their fitted magnitude are also shown separately. Top: \dphi correlation functions. Bottom: \dy correlation functions.}
\label{fig:fits1}
\end{figure}

In the next step, we test whether the LHCb data can constrain the different processes by fitting the data with the three individual \pythia templates while keeping the uncorrelated contribution fixed to the \pythia prediction. Since the \pT spectra are very similar for the three categories, the focus is purely on the correlations. Figure~\ref{fig:fits1} shows the results, where separate fits were performed for the \dphi and \dy distributions. In both fits, the preferred initial-state-shower contribution is very small. This suggests that the broad away-side component predicted by \pythia is not strongly favored by the data within the present template description. More differential measurements would help determine whether this reflects the relative contributions of the production mechanisms or differences between the simulated and measured correlation shapes. In the following, we therefore pursue a more direct experimental strategy: slicing the data in \dy to construct samples enriched in the different topologies and then examining their \dphi distributions. 

\begin{figure}
\centering
\includegraphics[width=0.9\linewidth]{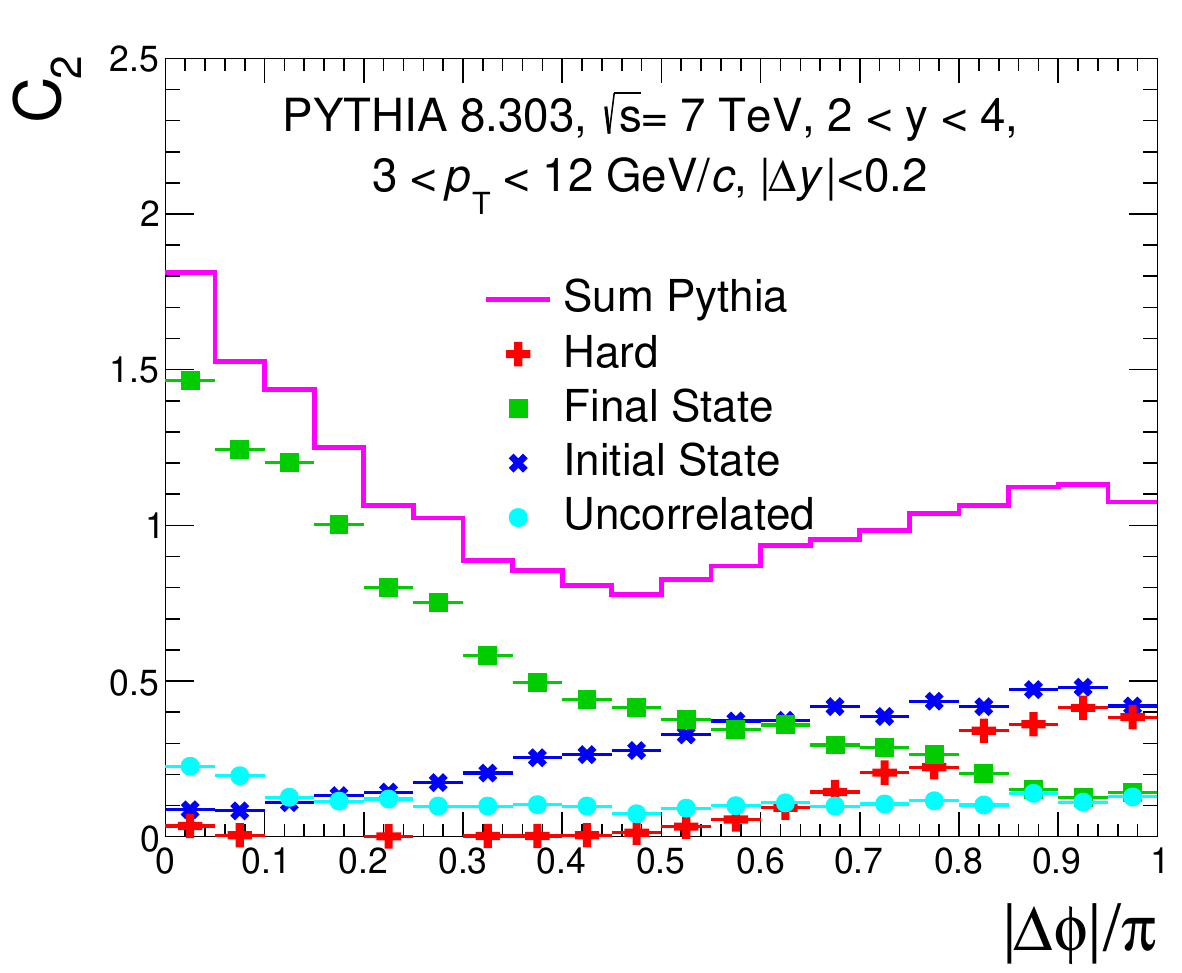}
\includegraphics[width=0.9\linewidth]{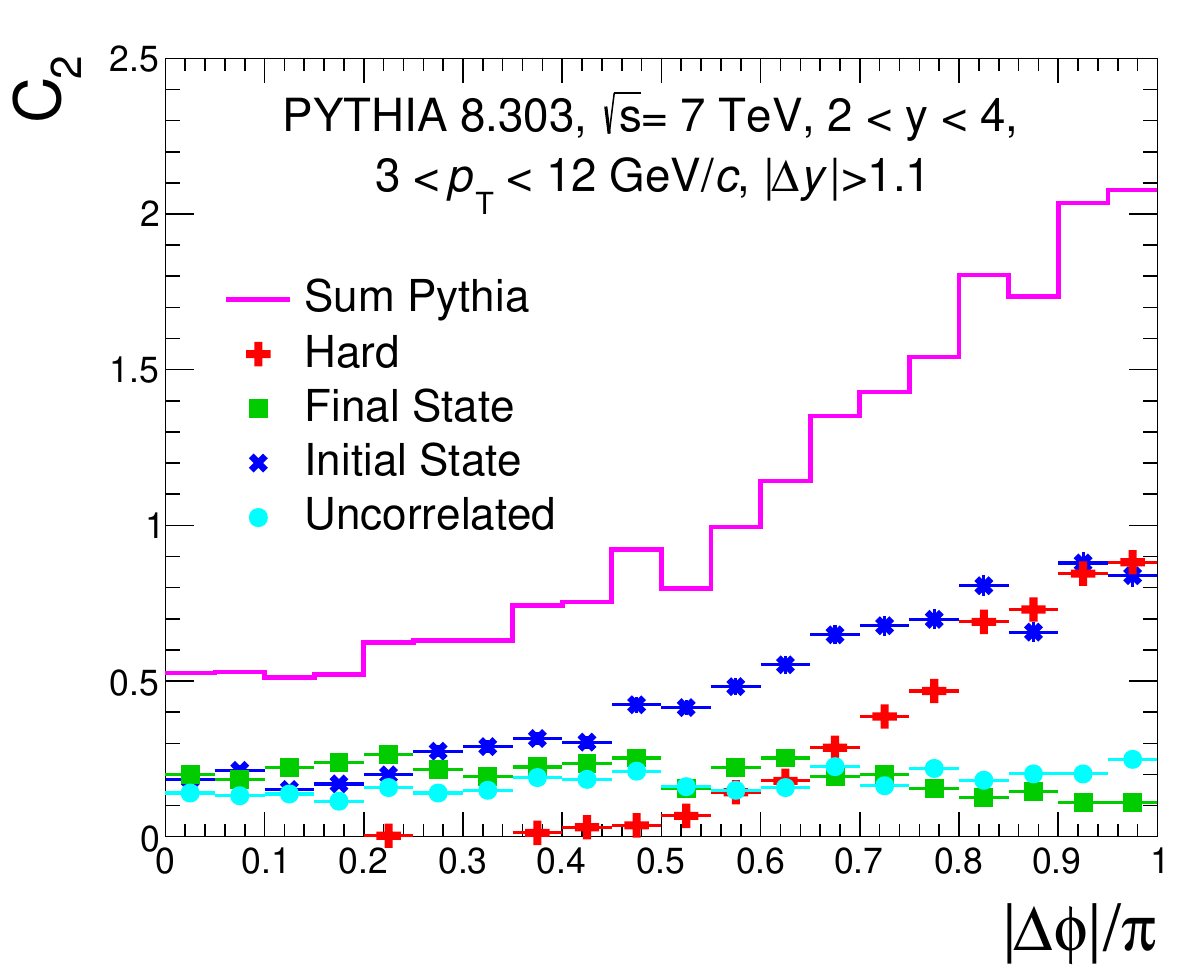}
    \caption{\dphi correlation functions from the \pythia simulations (pink lines) as well as the three event-history components when selections on $|\dy|$ are made for $3 < \pT < \gevc[12]$. Top: $|\dy| < 0.2$. Bottom:  $|\dy| > 1.1$.}
\label{fig:y_cuts1}
\end{figure}

Figure~\ref{fig:y_cuts1} shows that a narrow $|\dy|$ selection strongly enriches the final-state splitting component, while a broad $|\dy|$ selection suppresses it. The large-$|\dy|$ region therefore provides increased sensitivity to the hard-process and initial-state-shower categories, including the broad away-side component that does not appear to be well described by the default \pythia simulation. In this way one can therefore hope to understand more precisely if the component is as broad as \pythia predicts. 

\begin{figure}
\centering
\includegraphics[width=0.9\linewidth]{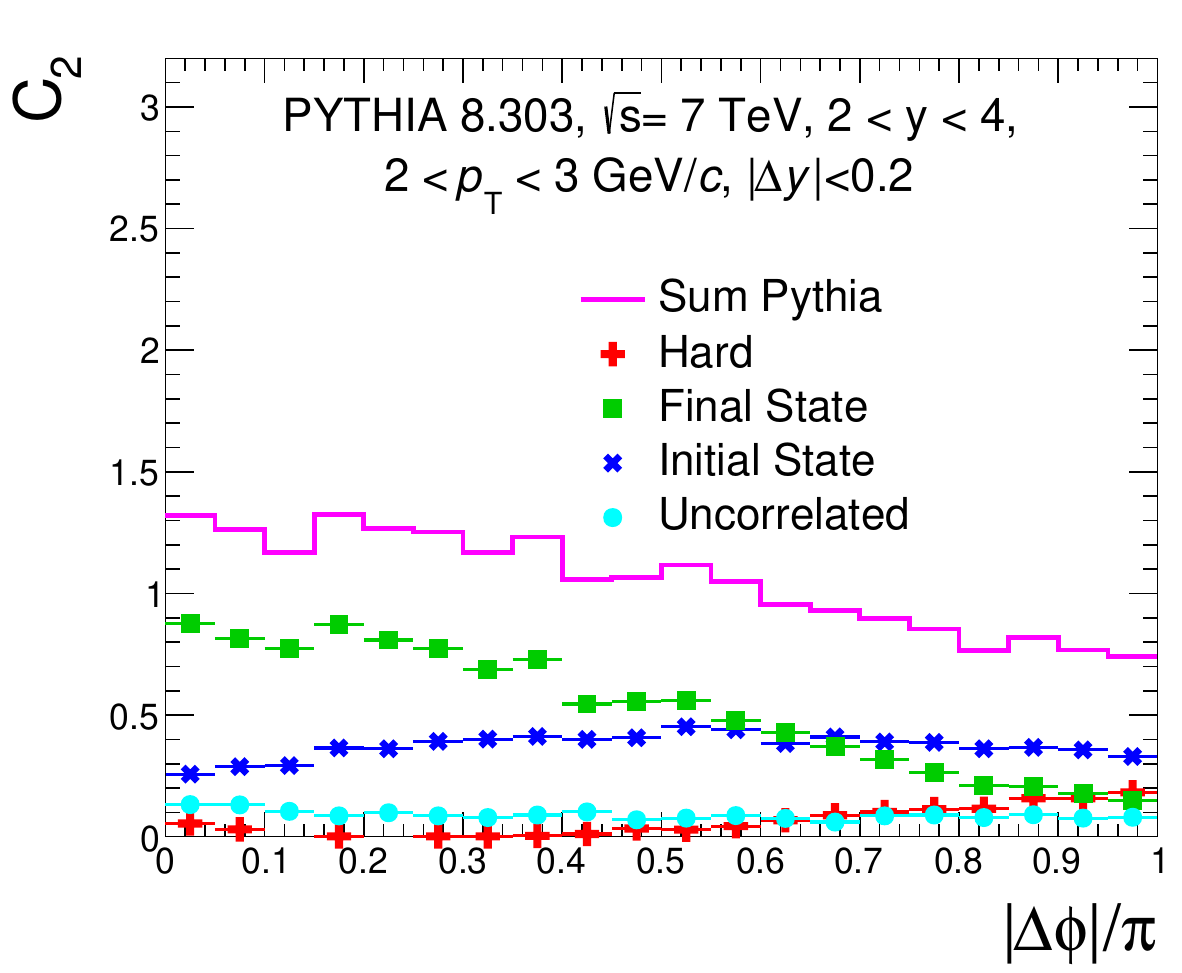}
\includegraphics[width=0.9\linewidth]{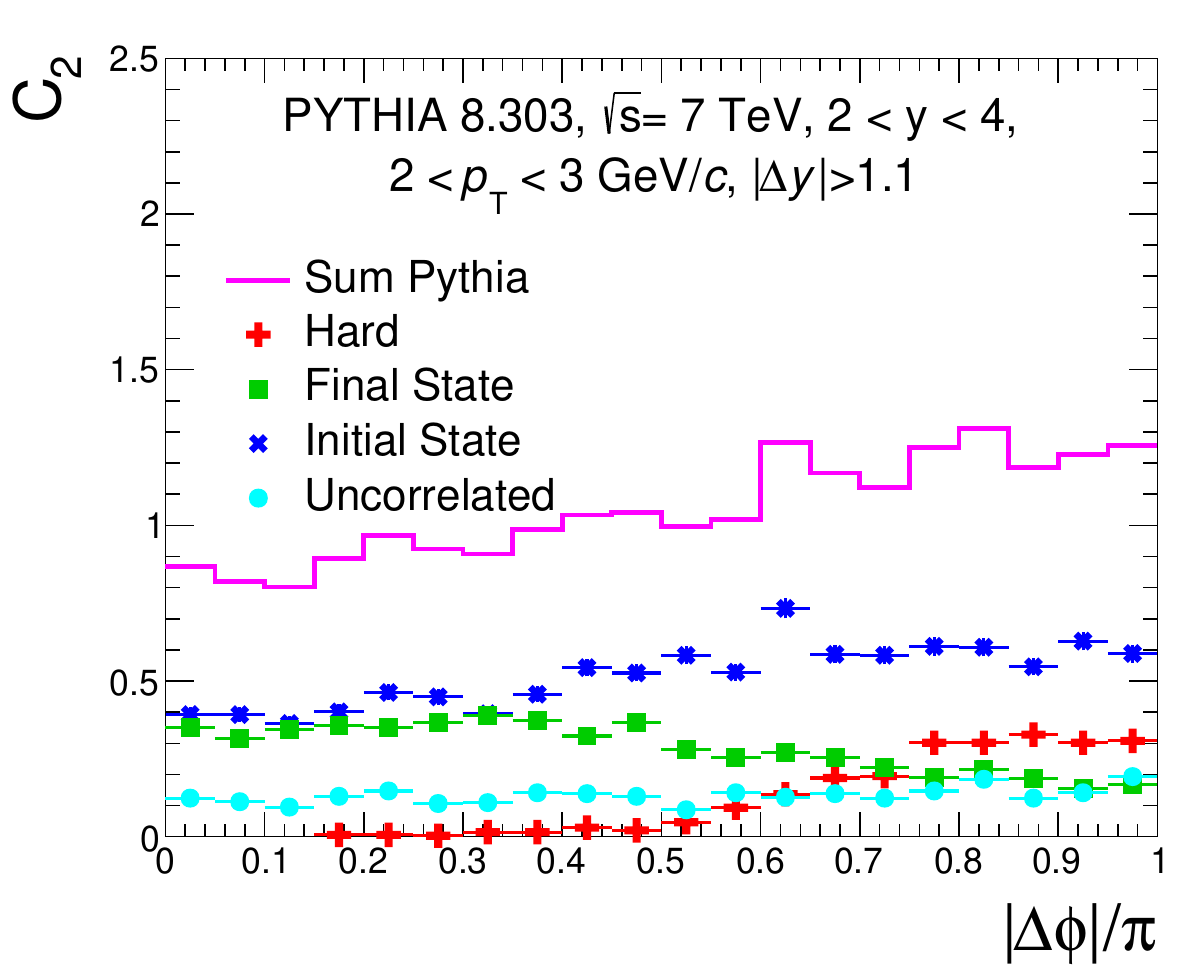}
    \caption{\dphi correlation functions from the \pythia simulations (pink lines) as well as the three event-history components when selections on $|\dy|$ are made for $2 < \pT < \gevc[3]$. Top: $|\dy| < 0.2$. Bottom:  $|\dy| > 1.1$.}
\label{fig:y_cuts2}
\end{figure}

\begin{figure}
\centering
\includegraphics[width=0.9\linewidth]{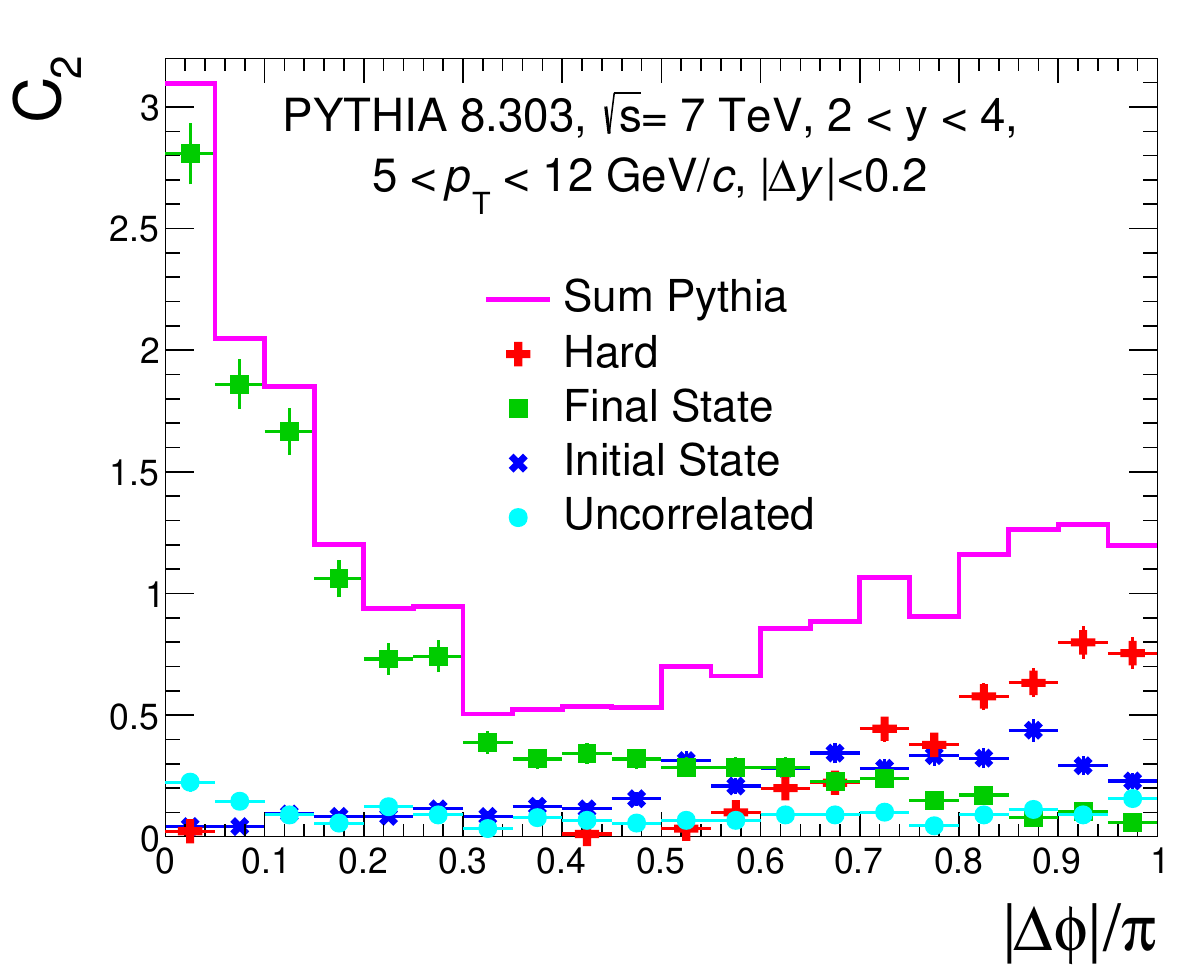}
\includegraphics[width=0.9\linewidth]{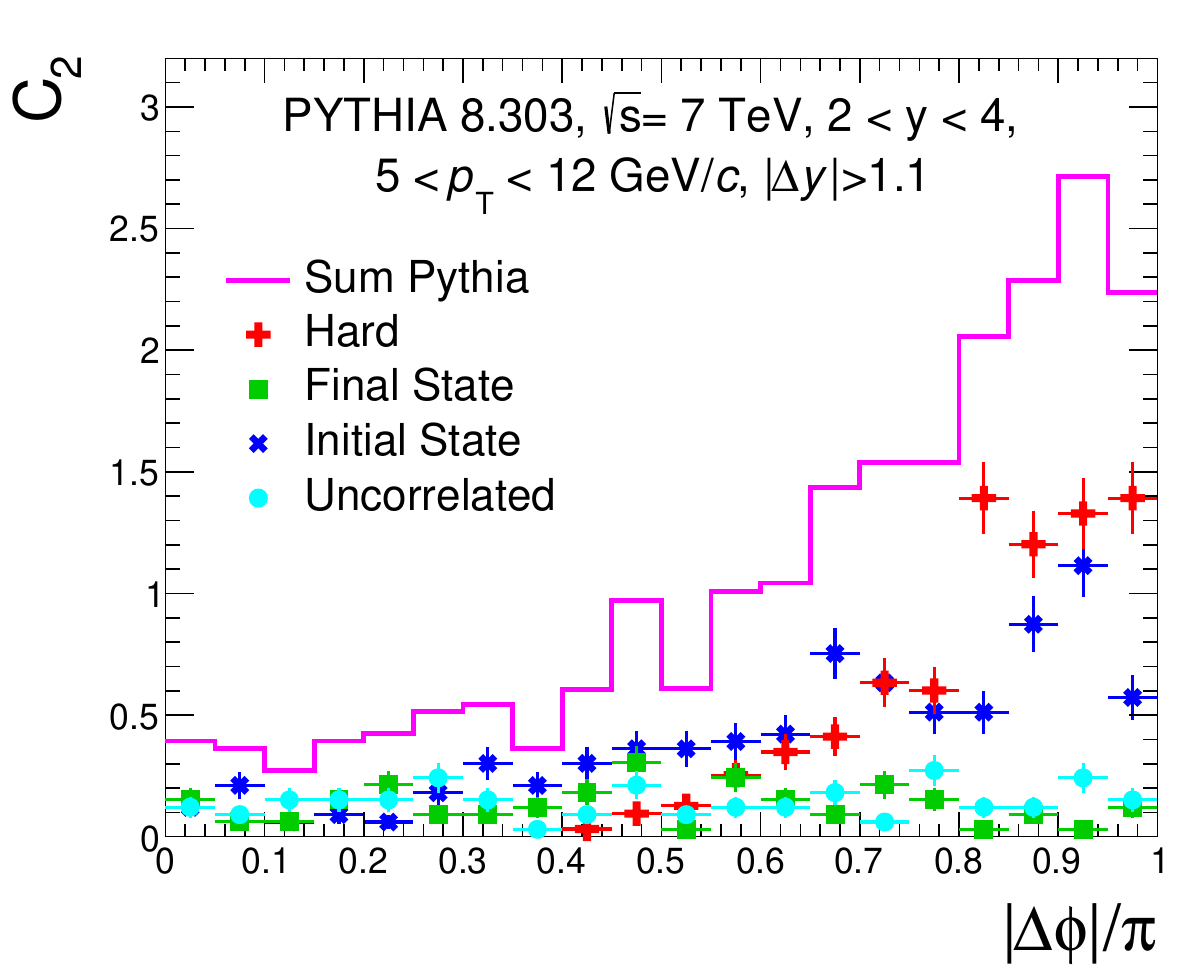}
    \caption{\dphi correlation functions from the \pythia simulations (pink lines) as well as the three event-history components when selections on $|\dy|$ are made for $5 < \pT < \gevc[12]$. Top: $|\dy| < 0.2$. Bottom:  $|\dy| > 1.1$.}
\label{fig:y_cuts3}
\end{figure}

In Fig.~\ref{fig:y_cuts2} and Fig.~\ref{fig:y_cuts3} we extend the idea to different \pT regions. Figure~\ref{fig:y_cuts2} shows that even in the interval $2 < \pT < \gevc[3]$ there is sensitivity to the final-state splitting and initial-state-shower contributions. This lower-\pT region is particularly interesting because it is also where the largest systematic uncertainties on pQCD predictions are observed. Finally, Fig.~\ref{fig:y_cuts3} shows that the separation between the different topologies becomes substantially sharper at higher \pT. This remains a region where systematic uncertainties of order 50\% can occur in pQCD calculations~\cite{ALICE:2023sgl}. This suggests that it might be easier to build the understanding starting from the higher \pT and then moving to lower \pT.

\section{Discussion}
\label{sec:discussion}

The very large amount of data collected at the LHC in Run 3 and to be collected in future runs motivates new differential observables that can improve our understanding of charm production. The previous section shows that selections on the rapidity difference of \ddbar pairs can construct samples strongly enriched in different production topologies. One could take this a step further and also select on \dphi to test whether the \pT spectra depend on the production topology. This is particularly promising at high \pT, where a narrow near-side $|\dphi|$ selection in Fig.~\ref{fig:y_cuts3} strongly enriches final-state splitting charm, while an away-side selection suppresses that contribution. One can find a similar idea in Ref.~\cite{Field:2002da} where it is proposed to use the jet-cone variable $R = \sqrt{\dy^2 + \dphi^2}$ to differentiate topologies. Related angular-separation observables have also been measured at the LHC, in particular to investigate gluon splitting into heavy-quark pairs~\cite{ATLAS:2018zhf}.

It is interesting to note that the fitted initial-state-shower component in Fig.~\ref{fig:fits1} is very small in both fits. This category includes the conventional flavor-excitation topology, but is defined operationally by the \pythia event history rather than by a unique decomposition of QCD amplitudes. We therefore interpret the small fitted contribution primarily as an indication that the broad away-side component predicted by the initial-state shower is particularly useful to constrain experimentally. Measurements differential in both \dy and \dphi could test how well this part of the event evolution is modeled.

\begin{figure}
\centering
\includegraphics[width=0.9\linewidth]{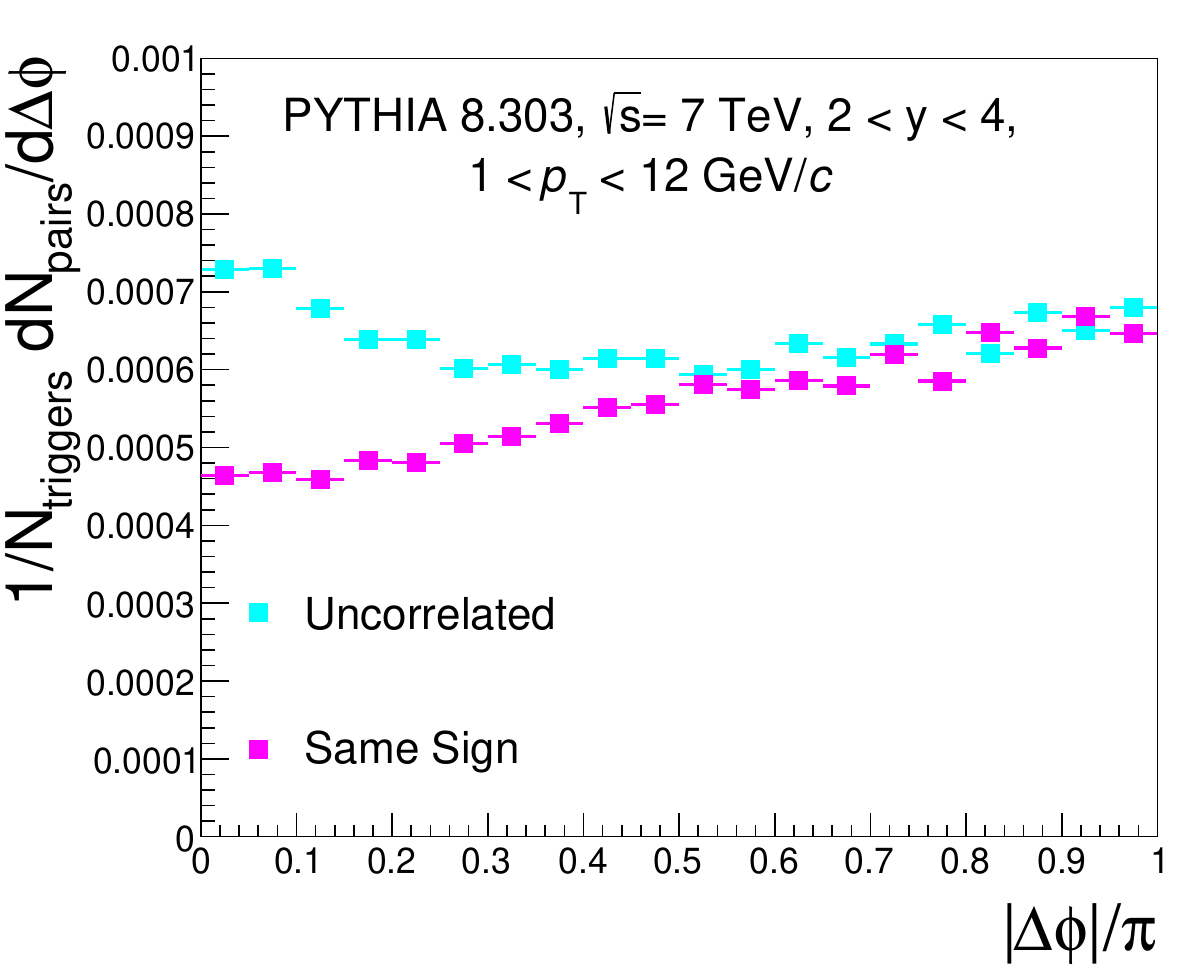}
\includegraphics[width=0.9\linewidth]{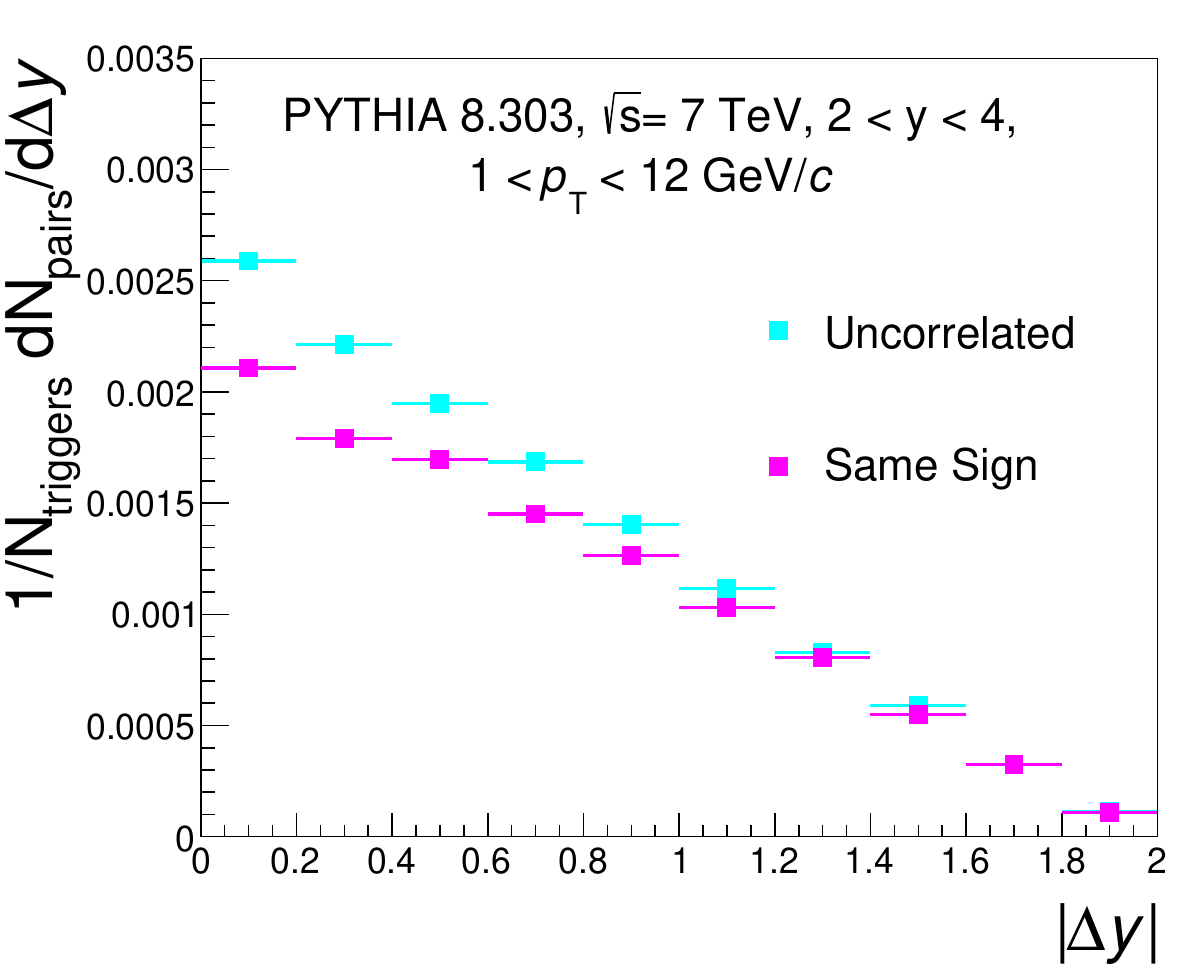}
    \caption{Per-trigger yields of uncorrelated \ddbar pairs compared with the per-trigger yields of same-sign pairs (\dd and \dbardbar). Top: \dphi per-trigger yields. Bottom: \dy per-trigger yields. }
\label{fig:balance}
\end{figure}

\begin{figure*}
\centering
\includegraphics[width=0.3\linewidth]{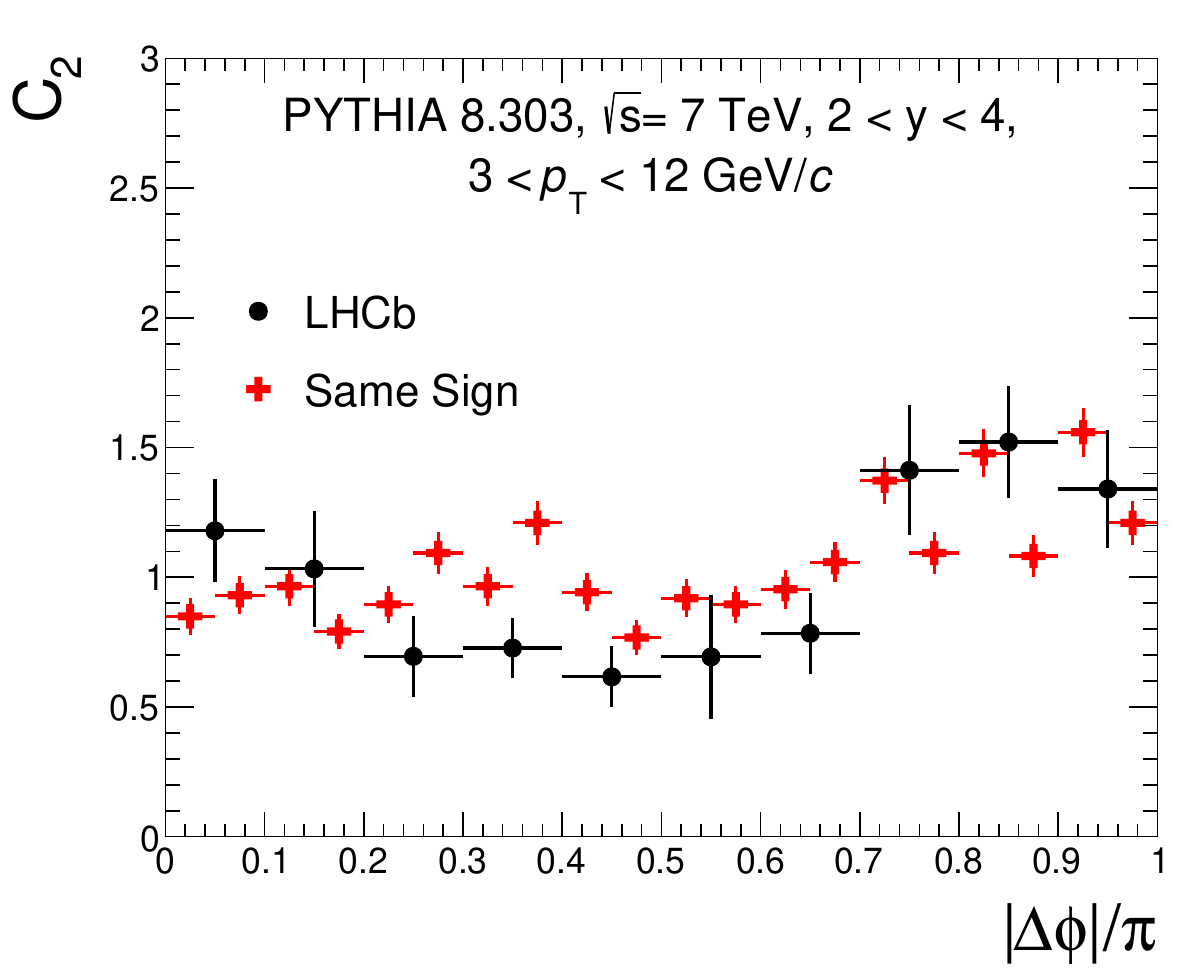}
\includegraphics[width=0.3\linewidth]{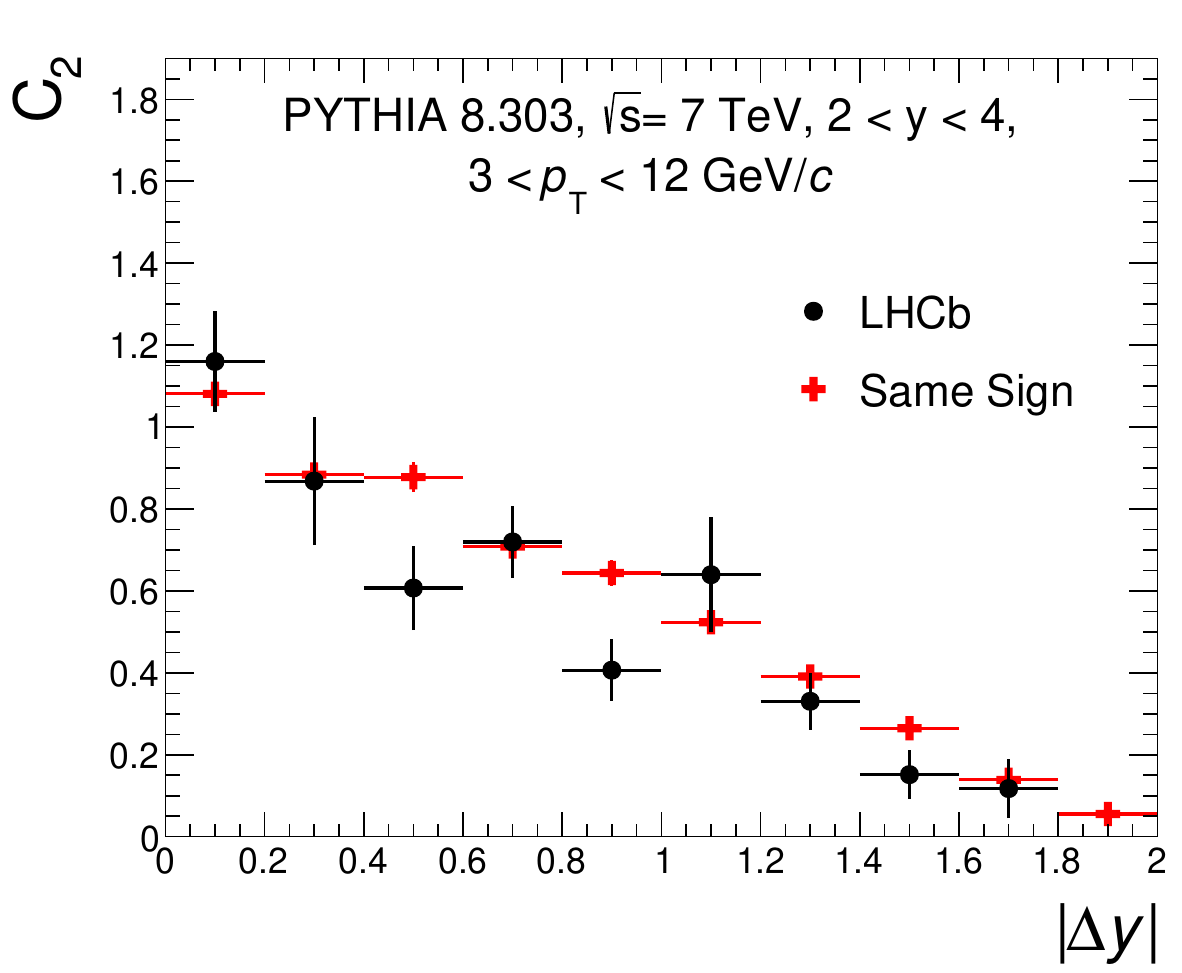}
\includegraphics[width=0.3\linewidth]{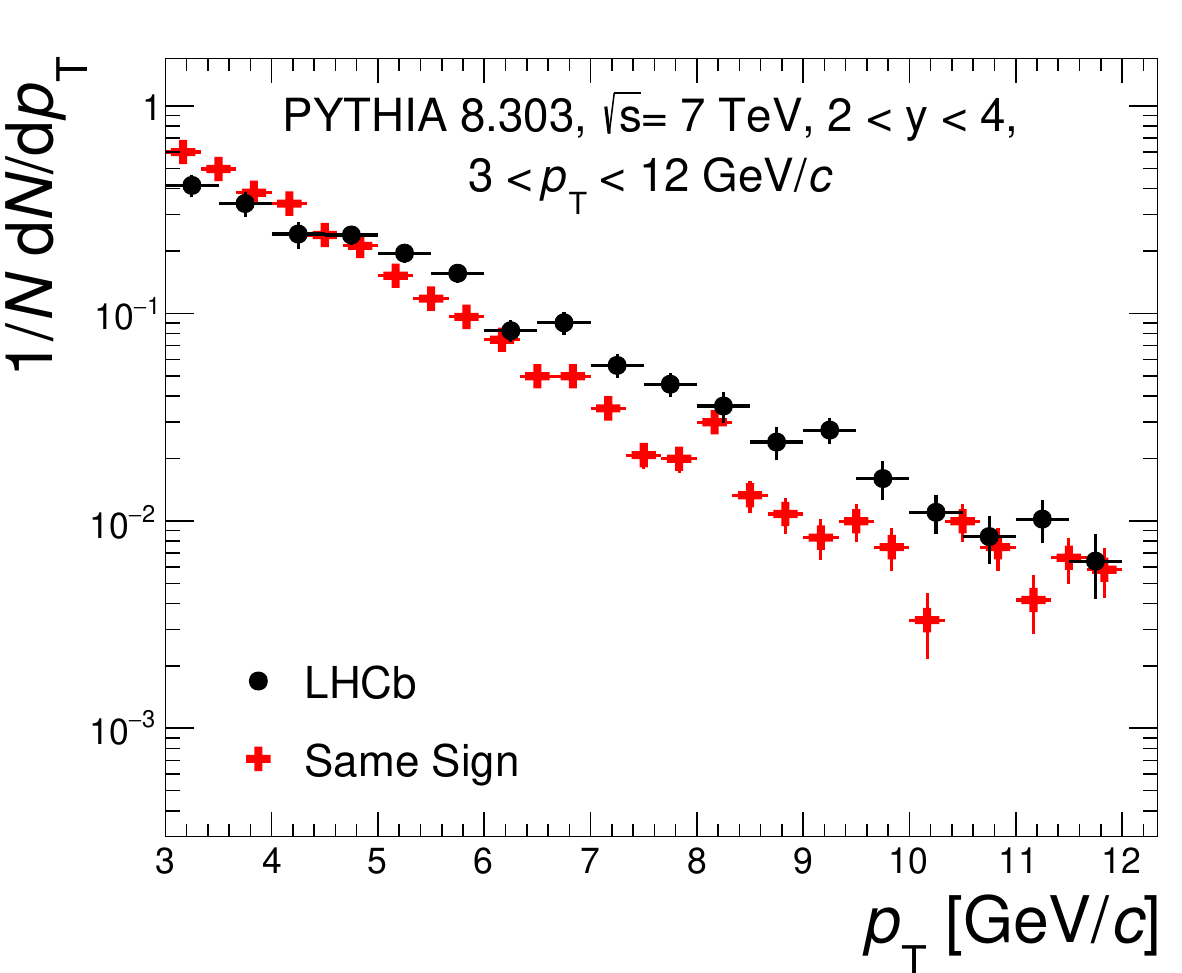}
    \caption{The distributions obtained with \pythia for \dd processes, compared with results from the LHCb experiment. Left: \dphi correlation functions. Center: \dy correlation functions. Right: \pT spectra. }
\label{fig:Same Sign}
\end{figure*}

Another possibility is, instead of correlation functions, to study per-trigger yields, $(1/N_{\text{triggers}})d^2N_{\text{pairs}}/d\dphi\dy$ where the trigger is either a $D^0$ or a $\bar{D}^0$ meson and the normalization is done to the number of triggers instead of the number of pairs. Correlation functions show only the shape (with a fixed integral), whereas per-trigger yields retain sensitivity to the actual number of triggers that is associated with a pair. This would also allow one to, at least in principle, estimate and subtract the uncorrelated background using same-sign correlations. We say ``in principle'' because we have tested this using \pythia and found that while they are similar they are not exactly the same. Figure~\ref{fig:balance} shows the comparison between the per-trigger yield of uncorrelated \ddbar pairs and that of \dd and \dbardbar. As can be seen, the distributions differ, which we understand is due to the following cluster-collapse effect in \pythia~\cite{Norrbin:1998bw}. If a collision produces two \ccbar pairs: $c_1\bar{c}_1$ and $c_2\bar{c}_2$, then the analysis of \ccbar pairs will identify four pairs (two correlated and two uncorrelated, $c_1\bar{c}_2$ and $c_2\bar{c}_1$). The analysis of same-sign pairs will find two pairs, $c_1c_2$ and $\bar{c}_1\bar{c}_2$, and so one would expect that they would be a good proxy for the uncorrelated pairs. However, as was shown in Ref.~\cite{Weber:2018ddv} it can happen in \pythia that an uncorrelated \ccbar pair forms a \jpsi particle. If this happens in the example with two \ccbar pairs then it leaves one uncorrelated pair but no same-sign pairs. This is also why same-sign pairs undershoot the uncorrelated pairs for $\dphi \sim 0$ and $\dy \sim 0$ in Fig.~\ref{fig:balance}. \\
Independently of this, we think it is still worthwhile to measure and report the per-trigger yield of uncorrelated pairs because it does provide a lower limit for the amount of uncorrelated pairs in the \ddbar sample. For completeness, we show in Fig.~\ref{fig:Same Sign} also a comparison of the correlation functions of \dd and \dbardbar measured by LHCb with the \pythia simulation. The \pythia simulations reproduce the main qualitative features of the measured distributions, although some differences remain, particularly in the \pT spectrum. More precise measurements, utilizing the higher statistics now available at the LHC, would provide useful constraints on the model. It is important to stress here that the \jpsi mechanism in \pythia has not been experimentally confirmed and so it would be worthwhile to see if one could test this. One can at least notice that the LHCb study also reported correlations between \jpsi and open charm hadrons and that these appear to be flat, albeit with low statistical precision, which one would not expect if the \pythia mechanism was important. More data might also make it easier to test this.

Before moving on to the final point, we just want to stress that this difference between uncorrelated opposite-sign pairs and same-sign pairs could also affect other measurements, e.g., of balance functions~\cite{Bass:2000az} or same-sign correlations between baryons where a dip is indeed observed~\cite{ALICE:2016jjg}. 

The final point we want to discuss is the relation to the investigation of the QGP. As mentioned in the introduction, charm is also an important probe of the plasma. The interaction of a parton in the plasma depends on its color charge and mass. This suggests that charm produced via gluon splitting might have a different sensitivity as it propagates through the medium because the ``parent'' gluon interacts stronger with the medium than the final two charm quarks. The characteristic formation length associated with gluon splitting scales as $E/m_{c\bar{c}}^2$ and so one expects this effect to mainly affect high \pT processes.
It would therefore be useful to construct enriched samples of hard-process charm and final-state splitting charm to test whether their modification in \pbpb collisions differs, in particular for jet quenching at high \pT. \\
Process-separated measurements could also be relevant for other scenarios, as it has, for example, been suggested that gluon splitting, $g \rightarrow \ccbar$, can be enhanced by the medium~\cite{Attems:2022otp}. 

ATLAS has measured azimuthal angle correlations (\dphi) between muons originating from semileptonic decays of heavy quarks produced in 5.02\,TeV
\pbpb and \pp collisions~\cite{ATLAS:2023vms}. The correlations are measured for muons with $\pT > \gevc[4]$ and $\eta < 2.4$ and the muon pairs are required to have $|\deta| > 0.8$ to reject \jpsi candidates. The study does not observe modifications in \pbpb collisions but also concludes, see supplementary material, that it is mainly sensitive to \bbbar pairs. If we take into account the lessons from the study presented here, then a large pseudorapidity separation preferentially suppresses the near-side final-state splitting topology and enhances sensitivity to away-side production. If one could perform an analogous study with, for example, \ddbar pairs, where the \jpsi rejection issue is different, selections in \dy could be used to construct samples enriched in hard-process and final-state splitting charm and test whether their medium modifications differ in \pbpb. Finally, one should ideally perform these measurements with per-trigger yields instead of correlation functions to be able to study how the yield is suppressed with respect to binary scaling in \pbpb collisions.

\section{Conclusions}
\label{sec:conclusions}

The \pythia study presented here demonstrates that different charm-production topologies lead to markedly different correlations in rapidity and azimuthal angle. Hard-process charm from flavor creation, $gg,q\bar{q}\rightarrow\ccbar$, is predominantly back-to-back and broad in \dy, while charm from final-state $g\rightarrow\ccbar$ splitting is concentrated at small \dphi and small \dy. Charm associated with the initial-state shower, which includes the conventional flavor-excitation topology, is also predominantly away-side but has a substantially broader azimuthal correlation. The comparison with existing LHCb \ddbar measurements illustrates the potential of such studies, but also indicates a need for more differential data to constrain the individual processes better. \pythia simulations indicate that selections on $|\dy|$ can strongly enrich samples in different production processes. The separation persists over a broad range in \pT and becomes particularly pronounced at high \pT.

Additional information could be obtained by measuring per-trigger pair yields rather than only normalized correlation functions, since this retains sensitivity to the absolute number of correlated and uncorrelated pairs, which appears to be important at low \pT.  

Finally, the same strategy could provide a new handle on charm production in heavy-ion collisions. Charm quarks produced directly in the hard scattering can probe the medium as heavy quarks from its earliest stages. In contrast, when a charm pair originates from final-state gluon splitting, the parent gluon may propagate through the medium and lose energy before splitting into the charm pair. Comparing samples enriched in these two production topologies could therefore reveal differences in their medium-induced modifications in \pbpb collisions. Such measurements offer the prospect of experimentally separating effects associated with the propagation of charm quarks from those associated with a parent gluon, providing a new probe of their different interactions with the quark-gluon plasma.

\section*{Acknowledgment}

The authors would like to thank Torbjörn Sjöstrand for useful suggestions during the development of this project and comments on the final manuscript.\\
Support from the following research grant is gratefully acknowledged: Vetenskapsrådet contract 2021-05179.

\bibliography{References}

\end{document}